# Gamma spectroscopic measurements using the PID350 pixelated CdTe radiation detector

K. Karafasoulis, K. Zachariadou, S. Seferlis, I. Papadakis, D. Loukas, C. Lambropoulos, C. Potiriadis

*Abstract–* Spectroscopic measurements are presented using the PID350 pixelated gamma radiation detectors. A high-speed data acquisition system has been developed in order to reduce the data loss during the data reading in case of a high flux of photons. A data analysis framework has been developed in order to improve the resolution of the acquired energy spectra, using specific calibration parameters for each PID350's pixel. Three PID350 detectors have been used to construct a stacked prototype system and spectroscopic measurements have been performed in order to test the ability of the prototype to localize radioactive sources.

## I. INTRODUCTION

THE PID350 provided by AJAT Oy is a pixelated gamma ray detector based on CdTe-CMOS technology (see Fig.1).

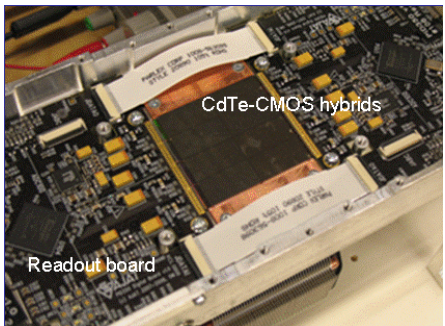

Fig. 1. The PID350 pixelated detector.

Manuscript received November 12, 2010. This work was supported by the collaborative Project COCAE SEC-218000 of the European Community's Seventh Framework Program.

K. Karafasoulis is with the Greek Atomic Energy Commission, Agia Paraskevi, Attiki, Greece and the Hellenic Army Academy, Vari, Attiki Greece (e-mail: ckaraf@gmail.com).

K. Zachariadou was with the Institute of Nuclear Physics, National Center for Scientific Research, Agia Paraskevi, Attiki, Greece. She is now with the Greek Atomic Energy Commission, Agia Paraskevi, Attiki, Greece and the Technological Educational Institute of Piraeus, Petrou Rali & Thivon-Athens, Greece (e-mail:zacharia@inp.demokritos.gr).

S. Seferlis is with the Greek Atomic Energy Commission, Agia Paraskevi, Attiki, Greece (e-mail: stsefer@eeae.gr).

I. Papadakis is with the Institute of Nuclear Physics, National Center for Scientific Research, Agia Paraskevi, Attiki, Greece (e-mail: papadakis@inp.demokritos.gr).

D. Loukas is with the Institute of Nuclear Physics, National Center for Scientific Research, Agia Paraskevi, Attiki, Greece (e-mail: loukas@inp.demokritos.gr).

C.P. Lambropoulos is with the Technological Educational Institute of Chalkida, Psahna – Evia, 34400 Greece (e-mail: lambrop@teihal.gr).

C. Potiriadis is with the Greek Atomic Energy Commission, Agia Paraskevi, Attiki, Greece (e-mail: cpot@eeae.gr).

The active area of the PID350 covers an area of 4.5 cm x 4.5 cm and consists of two modules. Each module has 8192 radiation sensing pixels with 350μm size, thus a PID350 detector consists of 16384 pixels.

The maximum energy deposited in each detector pixel, during two successive read out cycles, is stored into a specific address in the local PID350 memory. During one read out cycle, the contents of all the 8192 pixels of one PID350 module are read and stored in the computer, forming one frame.

The two PID350 modules are connected with two digital control boards having individual power supply and data readout (see Fig.2), thus the PID350 detector consists of two separate modules. Data are transferred from the control board of each module directly to a PC through a Serial Peripheral Interface (SPI) bus.

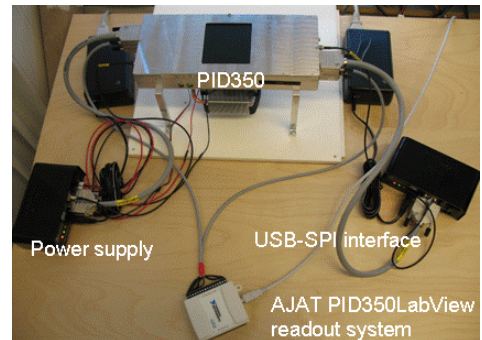

Fig. 2. The PID350 setup.

The PID350 standard data acquisition system has a maximum data rate transfer of 2 frames/s and has the ability to display the energy spectrum of each pixel, and the cumulative energy spectrum over all pixels.

Two different procedures are used for the energy calibration of the PID350. The first one, the hardware calibration, based on the offset and gain adjustment of each PID350 pixel, is supported by the standard setup provided by AJAT. The second one, the software calibration, is based on a software package, developed in the framework of the present study, in order to analyze data collected by a new faster data transfer system and to improve the energy resolution of the cumulative spectrum over all pixels.

In the present work three PID350 detectors have been used (labeled as PID350#1, PID350#2 and PID350#3. The two detector's modules are labeled with a subscript: e.g. the PID350#1 detector consists of the PID350#1_1 and PID350#1_2 modules) for spectroscopic measurements and their performance has been evaluated in order to be used in a stacked prototype system under development.

## II. HARDWARE CALIBRATION

The offset and gain adjustment of each PID350 module has been performed using a $^{241}$Am radioactive source.

For the offset adjustment, noise data have been collected with the PID350 standard acquisition system in order to locate the noise peak of each pixel. Several iterations are needed in order to reduce the width of the noise peak to no more than two channels..

For the gain adjustment, data have been collected for each PID350 detector using a $^{241}$Am radioactive source. The gain adjustment procedure searches for the gamma peak location and corrects the gain parameters in order to align the gamma peaks of all pixels. Several iterations have been performed in order to squeeze the distribution width of the gamma peak positions (centroids) to no more than two channels.

Since the offset and gain adjustments are not completely independent, they have been repeated iteratively several times, in order to reduce the FWHM of the cumulative gamma ray spectrum and to complete the hardware calibration.

The cumulative spectrum of $^{241}$Am radioactive source calibration is shown in Fig. 3 before and after the hardware calibration, for the case of the PID350#3 detector. Similar spectra have been obtained for the PID350#1 and PID350#2 detectors.

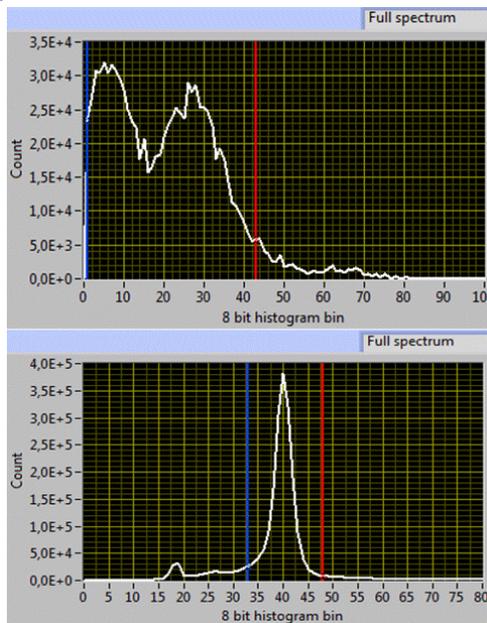

Fig. 3. Cumulative spectrum of $^{241}$Am radioactive source, with the PID350 detector, before (up) and after (down) the adjustment of the offset and gain parameters of all pixels.

## III. SOFTWARE CALIBRATION

In order to reduce the data loss during the data transfer (due to the slower read out cycle from a PID350 to the computer, compared to the internal memory writing speed) as well as to improve the energy resolution of the cumulative energy spectrum collected by the PID350 modules a) a high speed readout system and b) a data analysis framework have been developed.

### A. High Speed Readout System

The high speed readout system developed to reduce the data loss consists of an FPGA SPI, a High Speed USB, and user interface software written in VHDL and C languages.

The system (see Fig. 4) increases the data transfer speed to the PC from 2 frames/s, which was the data transfer speed of the standard PID350 readout system, up to 30 frames/s.

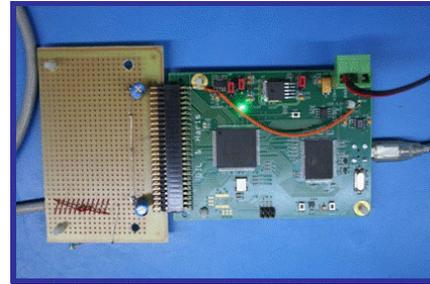

Fig. 4. The high speed PID350 data acquisition system.

The user interface software of the high speed readout system checks the PID350 detectors status and temperature, uploads the offset and gain parameters derived by the hardware calibration procedure described in the previous section and starts the data gathering.

### B. Data Analysis Framework

The raw data recorded with the high speed readout system are stored in binary data files in a stream of bytes which are grouped in 8-byte packets. Each packet stores the content of a single pixel from every PID350 module. This leads to 112.5 Mbytes/min. In order to manipulate the PID350 raw data with greater flexibility a set of C++ classes has been developed under the ROOT framework [1], that transforms the raw data packets into usable frames (a frame holds the contents of all pixels of a single PID350 module) and stores them in ROOT Tree format. Then the functionality of the ROOT framework can be used for analyzing further the data.

The data collected are stored in frames, where each frame holds the following information: an identification number for the frame (ID), the signal amplitude collected by each pixel, the spatial coordinates of each pixel as well as a sequence number which carries an estimation of the readout time. Furthermore, the software classes incorporated into the ROOT check the spectroscopic quality of each pixel applying the quality criteria defined in the paragraph V.

### C. Software energy calibration

Although the width of the distribution of the photo-peak centroids has been adjusted to be no more than two bins during the hardware calibration by using a $^{241}$Am source (as described in section II), it is much wider when a different mono-energetic photon field is used. This is evident in Fig. 5 that illustrates the photo-peak centroid distribution of all

PID350#3 pixels (the bad pixels are excluded), irradiated by a [109]Tc radioactive source.

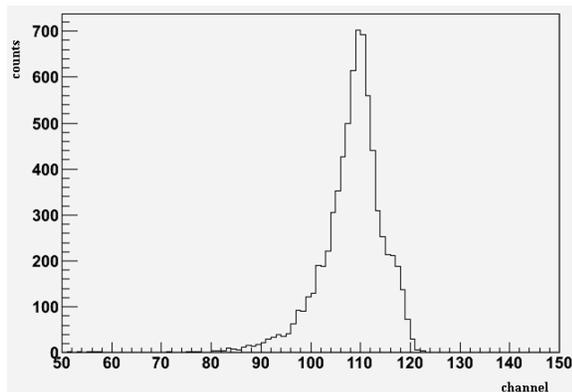

Fig. 5. Photo-peak centroid distribution for the PID350#3 detector illuminated with a [99m]Tc radioactive source.

The broad distribution of the photo-peak centroid results to a broad energy peak in the full energy spectrum, since the energy peak is the convolution of the width of each pixel's photo-peak with the distribution of the photo-peak centroids, as is shown in Fig. 6-a. To improve the energy resolution of the cumulative spectra a new calibration procedure is introduced.

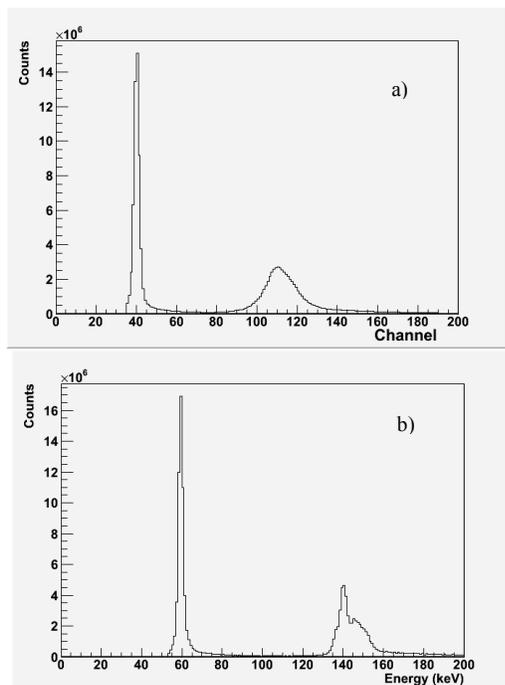

Fig. 6. The cumulative spectrum of a [241]Am and a [99m]Tc radioactive source recorded with the PID350#3 detector, (a) after hardware calibration and (b) after the software calibration procedure.

During this procedure two calibration constants are determined for every single pixel of the PID350 detector by using a [241]Am (60keV) and a [99m]Tc (140.5 keV) radioactive source. The energy spectrum after the software calibration, for the PID350#3 detector is shown in Fig. 6-b.

## IV. SIMULATION STUDIES

The complete PID350 detector geometry has been simulated using the GEANT package [2] with the aid of the open-source object oriented software library (MEGAlib [3]).

During the simulation studies a large number of photons ($10^9$) emitted from a point isotropic source placed on its axis of symmetry at a distance of 2.5 cm from the external surface of the PID350 and interacted with the PID350 detector model. The deposited energy was smeared using a Gaussian distribution with FWHM equal to 7 keV [4].

Fig. 7 presents the simulated energy deposition of all hits for incident gamma rays of 60 keV. The photo-peak (59.5 keV), the Ka x-ray from Cd (23 keV) and the escape peak (26.5 keV), broadened due to the smearing, are visible.

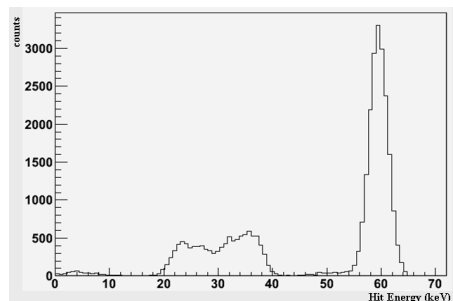

Fig.7. Simulated spectrum of the energy deposited in each PID350 detector pixel, for 60 keV incident gamma rays.

## V. THE PID350 STACKED PROTOTYPE SYSTEM

A prototype system has been assembled as a stack of three PID350 detectors stacked together. The experimental setup used to test the performance of the prototype system is shown in Fig.8.

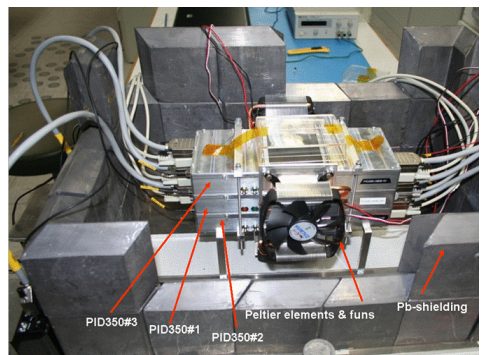

Fig.8. Experimental setup with the PID350 stacked prototype.

The ordering of the PID350 layers in the stacked prototype reflects the performance of each layer, i.e. the PID350s are stacked from the top to bottom with decreasing quality. The quality is defined by three parameters: the energy resolution, the number of bad pixels and the upper channel limit of the noise peak.

To determine the energy resolution of each PID350 detector, data have been collected using a [241]Am radioactive source. The calibrated photo-peaks of each PID350 are illustrated in Fig. 9.

(b)

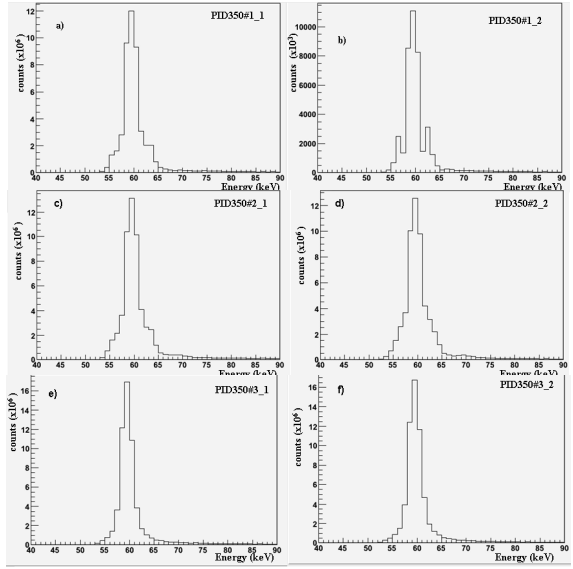

Fig. 9. Calibrated photo-peaks of $^{241}$Am radioactive source acquired with a) PID350#1_1, b) PID350#1_2, c) PID350#2_1, d) PID350#2_2 e) PID350#3_1 and f) PID350#3_2 detector modules.

The noise distribution of each PID350 detector is shown in Fig. 10. The detectors have sharp noise distributions except for PID350#2 which has a long tail to the right.

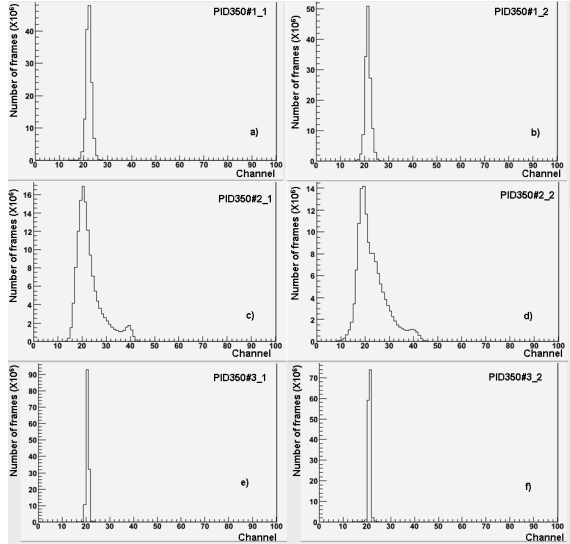

Fig. 10. Noise distribution for a) PID350#1_1, b) PID350#1_2, c) PID350#2_1, d) PID350#2_2 e) PID350#3_1 and f) PID350#3_2 detector modules.

Finally, the map of the bad pixels of each PID350 is shown in Fig. 11. The bad pixels are defined as the pixels for which the energy calibration procedure fails (the calibration algorithm can't find two peaks in order to calibrate the pixel), or the pixel is noisy (it has noisy channels above the upper edge of the dc level).

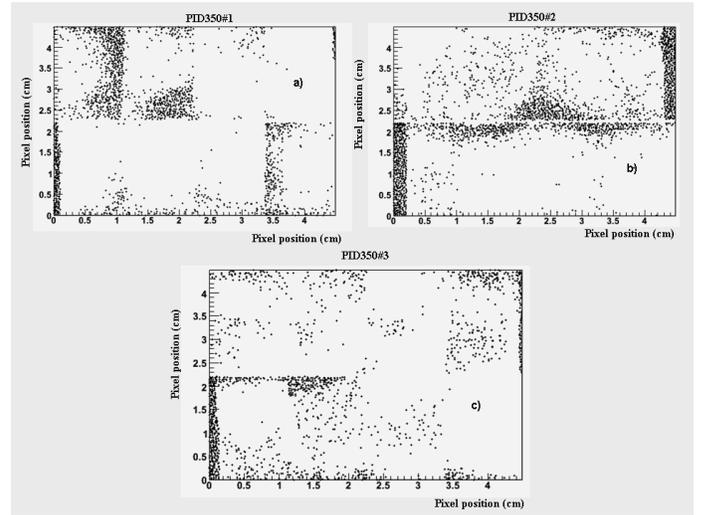

Fig. 11. Map of the bad pixels for the a) PID350#1, b) PID350#2 and c) PID350#3 detectors.

The quality parameters of each PID350 detector are summarized in Table I.

TABLE I. QUALITY PARAMETERS OF THE PID350 DETECTORS

| PID350 deetector | Noise channel | Energy Resolution σ(keV) | Bad pixels |
|---|---|---|---|
| PID350#1 | 26 | 1.73 | 1765 |
| PID350#2 | 44 | 2.15 | 2567 |
| PID350#3 | 22 | 1.74 | 1488 |

In order to perform Compton reconstruction [5] using the stacked prototype system, all the pixels activated by a photon interaction must be grouped together to form a Compton event.

Taking into account the time of a PID350' module data-reading (of the order of 30ms), it is expected that pixels activated by different incident photons will be recorded in the same data frame (pile-up). In order to reduce the photon flux, the $^{99m}$Tc (140.5 keV) radioactive source was collimated using a lead block having an opening of 1mm diameter. The source was placed was placed in front of the stacked detector, at a distance of 14 mm from its top layer and on its axis of symmetry. The radius of beam spot on the first layer is estimated to be 8 mm.). The distribution of the number of the activated pixels per frame recorded by one module of the first layer is presented in Fig. 12. The pixels with full energy deposition have been excluded from the distribution. The mean number of activated pixels per frame is 96, much higher than the mean number of activated pixels expected by a single incident photon. This fact implies that multiple incident photons are still recorded in each frame prohibiting the Compton reconstruction.[1]

---

[1] The PID350 stacked prototype is used as a precursor setup of the COCAE detector under development within the framework of the COCAE project. (http://www.cocae.eu) In the COCAE detector special care has been taken in order to overcome the problem of pile-up.

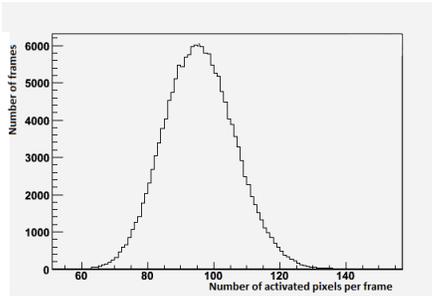

Fig. 12. Distribution of the number of activated pixels in the PID350#3_1 detector of the stacked system.

## VI. Source-detector Distance Estimation

In order to test the ability of the stacked detector to estimate the distance of a radioactive source, it has been irradiated by a $^{99m}$Tc (140.5 keV) source placed at various distances from the system's upper detecting layer. The source-to-detector distance estimation has been based on the number of the fully absorbed photons in each detecting layer.

The number of photo-peak counts ($N_i$) from each PID350 detector layer (i), is evaluated by fitting the function:

$$N_i \propto \exp\left(-(i-1)\left(\left(\sum_j \mu_j t_j\right) + a\right)\right) \cdot \left(\frac{z}{z+(i-1)g}\right)^2 \quad (1)$$

where $t_j$ is the thickness of a material of each detecting layer with corresponding total absorption coefficient $\mu_j$, g is the distance between the layers, α is a parameter evaluated experimentally and z the source-to-detector distance determined by the fit. The sum runs over all materials of the i$^{th}$ layer.

The result of the experiments for the z determination using the $^{99m}$Tc radioactive source is presented in Fig. 13. It can be noticed that the 3-layer PID350 stacked prototype system is capable of evaluating the distance of a gamma ray source with good accuracy place in the distance range from 20 cm up to 100 cm.

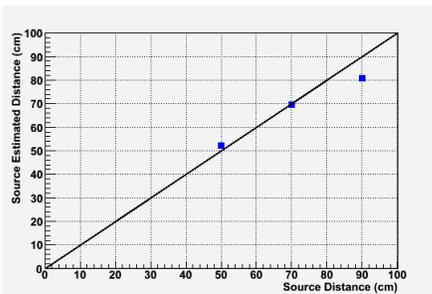

Fig. 13. Estimated distance vs. real distance of a Tc-99 radioactive source using (1).

## VII. Conclusion

Spectroscopic measurements using PID350 gamma radiation pixelated detectors are reported.

The hardware calibration (based on offset and gain adjustment) of each PID350 detector has been performed using the standard PID350 data acquisition system.

The complete PID350 geometry has been modeled and Monte Carlo simulations were performed with a large number of 60keV photons, interacting with the PID350 detector model. The simulated energy spectrum is presented and interpreted.

In order to reduce the data loss in case of a high flux of photons, a data acquisition system has been developed. The system increases the data transfer speed from 2 frames/sec which was the data transfer speed of the standard PID350's readout system up to 30 frames/sec.

Furthermore, a data analysis framework has been developed that transforms the raw data collected by the PID350 detectors into useable frames. Within this framework, an algorithm has been developed that performs software energy calibration for each detector pixel independently and improves the PID350 energy resolution.

Three PID350 detectors have been assembled in a stacked prototype system. Although the localization of a source by exploiting the Compton technique has been proven to be difficult, the system is capable of accurately evaluate the distance of radioactive sources, using the photo-peak count information from each layer.